%Paper: cond-mat/9512010
%From: mstone@esi.ac.at (Michael Stone)
%Date: Sat, 02 Dec 1995 12:24:50 +0100 (MET)

%%%%%%%%%%%%%%%%%%%%%%%%%%FORMAT%%%%%%%%%%%%%%%%%%%%%%%%%%%%%%%%%%%%%%%
\baselineskip=20pt
\parindent 20pt
\settabs 4 \columns
\parskip=10pt
\hfuzz=2pt   %accept some black boxes as ok
%%%%%%%%%%%%%%%%%%%%%%%%%MYDEFS%%%%%%%%%%%%%%%%%%%%%%%%%%%%%%%%%%%%%%%%%%%%
\def\line{\hbox to \hsize}
\def\frac #1#2{{#1\over #2}}

\def\A{{\cal A}}

\def\F{{\cal F}}
\def\J{{\cal J}}
\def\Z{{\cal Z}}
\def\N{{\cal N}}

\def \n{{\bf n}}
\def \r{{\bf r}}
\def \R{{\bf R}}

\def \z{{\bar z}}

%%%%%%%%%%%%%%%%%%%%%%%%%%%%%%%%%%%%%%%%%%%%%%%%%%%%%%%%%%%%%%%%%%%%%%%%%%

\line{\hfil }
\line{\hfil }
\line{\hfil November 1995}
\vskip 1cm
\centerline{\bf THE MAGNUS FORCE ON SKYRMIONS IN FERROMAGNETS }
\centerline{\bf   AND QUANTUM HALL SYSTEMS }
\vskip 1cm

\centerline{Michael Stone}
\centerline{\it Department of Physics}
\centerline{\it University of Illinois at Urbana Champaign}
\centerline{\it 1110 W. Green St.}
\centerline{\it Urbana, IL 61801}
\centerline{USA}

\line{\bf Abstract\hfil}

The topological solitons, or ``skyrmions'',
in a planar ferromagnet experience a Magnus force proportional to the
product of their   velocity and the surrounding magnetization.  It
has been suggested that the  charged quasiparticles  near filling
factor $\nu=1$ in the $GaAs$  quantum Hall effect  are  skyrmions. If
so, we might expect this spin-induced Magnus force to act on the
quasiparticles in addition to the Lorentz force  they experience
because of their charge. We show that this is not the case, and that the
Magnus and Lorentz  forces are merely different descriptions  of the
same physical effect.

\vfil

\eject

\line{{\bf 1. Introduction}\hfil}

Sondhi {\it et al.\/}  have argued [1] that the lowest energy  charged
quasiparticles  near filling fraction $\nu=1$ in the $GaAs$ quantum
Hall effect are  topological solitons, or ``skyrmions''.
The $\nu=1 $ ground-state is a filled spin-polarized
Landau level. For non-interacting electrons the elementary
excitations consist of the addition of  a single electron with
reversed spin, or the removal of a single electron from the fully
polarized Landau level.  The change in spin from the addition or
removal of an electron is in each case $\Delta S_z=-\frac 12$.
Skyrmions are different. They consist of an extended  region where
the spin direction gradually twists. This slowly varying spin texture
serves to bind or repel a unit charge, so the skyrmions still  have
charge $q=\pm e$,  but their total spin is much larger than $\frac 12$.
They are not perturbatively related to the single-particle elementary
excitations. Skyrmions will be energetically favored over the
$S_z=\frac 12$ quasiparticles whenever sharply localized charge
 fluctuations require more energy than overturning a number of spins
 ---{\it i.e.\/} when the gyromagnetic ratio $g$ is small [1].

The skyrmion picture has received strong support from recent
measurements   by Barrett {\it et al.\/} [2]. These authors see a
precipitous fall in the spin polarization of the electron gas on both
sides of $\nu=1$.  This clearly indicates that many spins are being
overturned by the addition or removal of a single electron.

 By looking for motional narrowing effects it is possible that the
NMR methods used in [2] could  probe the mobility of
the quasiparticles. Before attempting to calculate the mobility by
taking into account dissipative effects and quantum diffusion
 it is however necessary to have a thorough understanding of the
 quasi-classical forces acting on a moving skyrmion.
Now the  skyrmion configurations in a conventional ferromagnet
experience a ``Magnus'' force proportional to the product of their
velocity and the local magnetization. This force prevents the skyrmion
from moving with respect to the spin background. Because the skyrmion
quasiparticles in the quantum Hall effect are electrically charged we
would  expect them to experience a  Lorentz force in addition to the
Magnus force. This Lorentz force is of the same magnitude as the
Magnus force and one might hope that the two forces would cancel,
allowing the skyrmions to move freely. Sadly this does not happen. It
is easy to see that the two forces are {\it identical\/} in both
magnitude and sign. This equality is no coincidence. There is really
only {\it one} force, and its two apparently distinct origins are
  merely differing interpretations of a single geometric phase.
Only one of the two forces should therefore be taken into
account when considering the motion of the skyrmions. The present
paper is devoted to a discussion of this.

In section two we review the lagrangian approach to the dynamics of a
spin. We then show how the Magnus force appears in a conventional
ferromagnet composed of neutral spins obeying the Landau-Lifshitz
equation. In section three we show how  Landau-Lifshitz  dynamics
arises in a simple model for the Hall effect. We then use a duality
transformation to demonstrate that the Lorentz force is merely the
Magnus force in disguise. The last section provides a simple physical
explanation of why there is only one force.

\vskip 12pt
\line{{\bf 2. Ferromagnets} \hfil}

To establish our notation we start with a brief review of the
dynamics of a single spin, whose direction we denote by the unit
vector $\n$. The classical action for a spin in a magnetic field
${\bf B}$ is  a functional of the spin trajectory, or history,
$\n(t)$ and is given by [3,4]
$$
S=-J\int \dot \n \cdot {\bf A}(\n)\,dt+\mu\int {\bf B}\cdot \n\,dt.
\eqno(2.1)
$$
The second term in this expression is simply $-\int H\,dt$ where
$H=-\mu\, {\bf B}\cdot \n$ is the hamiltonian of for a  spin of moment
$\mu\, \n$ in the field ${\bf B}$.
The first term is more complicated. Here ${\bf A}(\n)$ denotes the
gauge potential of a unit (flux $=4\pi$) ``magnetic'' monopole located
at the center of the unit sphere $S^2$ on which $\n$ lies. For open
trajectories this term depends on the particular gauge chosen for
${\bf A}$, but  when the
motion of the  $\n$ vector is required to be periodic, as for example
when computing partition functions, we can rewrite (2.1) in a
manifestly gauge invariant manner as [5]
$$
S=-J\int\negthinspace\negthinspace\int
\n\cdot(\partial_\tau\n\wedge\partial_t\n)\,d\tau\, dt
+\mu\oint {\bf B}\cdot\n\, dt.
\eqno(2.2)
$$
The coordinates $t$ and $0\le\tau\le 1$ parameterize the interior of
the region  $\Gamma \subset S^2$ bounded by the curve $\n(t)$ on
which $\tau=1$. We have extended the definition of the function
$\n(t)$ to a function $\n(t,\tau):\Gamma\to S^2$ such that
$\n(t,1)=n(t)$. The numerical value of the action $S$ is independent
of the choice of extension. Geometrically, the first term in (2.2) is
simply the oriented area  of the  region $\Gamma$.

The classical equation of motion is found by varying $\n(t)$ in (2.2).
The variation of the first term is most easily obtained from its
geometric  interpretation. We evaluate the change in the area of
$\Gamma$ due to the variation in its boundary and so find
$$
\delta S=- J\oint (\n\cdot (\delta \n \wedge \dot \n))\,dt +
\mu\oint {\bf B}\cdot \delta \n \,dt.
\eqno (2.3)
$$
In order to respect the constraint on  the length of $\n$ we can write
$\delta\n=\n \wedge\delta {\bf w}$, whence
$$
\delta S=\oint  \delta {\bf w}\cdot
(J\dot \n- \mu(\n\wedge {\bf B}))\,dt.
\eqno (2.4)
$$
The action principle therefore tells us that
$$
J\dot\n-\mu(\n\wedge {\bf B})=0.
\eqno (2.5)
$$
We see that the spin undergoes the desired Larmor precession about the
direction of the $B$ field.

 The role of the ``monopole'' is made clear by acting
on (2.5) with $\n\,\wedge$ to get
$$
J(\dot \n \wedge \n)+\mu({\bf B}-({\bf B}\cdot \n )\n)=0.
\eqno(2.6)
$$
The first term in  (2.6) can now be interpreted as the ``Lorentz''
force on a  particle of ``charge'' $J$ constrained to move on $S^2$
in the field of the monopole. The other term, the  component  of
$\mu\,{\bf B}$ tangent to the sphere,  is the force attempting to align
the spin along  the direction of ${\bf B}$. The particle is massless
so the two forces acting on it must add to zero.

When the system is quantized by placing the classical action  in the
exponent of a path integral $\Z=\int d[\n]\exp i S$, the ambiguity
in the region (``inside'' {\it vs\/}, ``outside'') bounded by $\n(t)$
requires $J$ to take integer or half-integer values, giving rise to
the familiar quantization of angular momentum. The quantization of
spin by this method is part of the general theory of group
representations via the method of co-adjoint orbits [6].

We can immediately extend these ideas to a continuum model of a
ferromagnet with $\rho$ spins per unit area,
each  of magnitude $J$.  The only modifications required are  to make
$\n$ a function of position as well as time, and to replace the
external ${\bf B}$ field by  a spin-stiffness term. We therefore take
$$
S=-J\rho\int  \dot
\n \cdot {\bf A}(\n)\, dt\,d^2x-\frac 12 K \int (\nabla\n)^2\, dt\,d^2x,
\eqno(2.7)
$$
or equivalently, for periodic histories,
$$
S=-J\rho\int(\n\cdot (\partial_\tau\n \wedge \partial_t\n))\,d\tau\,
dt\,d^2x-
\frac 12 K \int (\nabla\n)^2\,dt\,d^2x.
\eqno (2.8)
$$
The corresponding equation of motion,
$$
J\rho\,\dot \n -K \n\wedge \nabla^2\n=0,
\eqno(2.9)
$$
is the Landau-Lifshitz equation [7] describing the precession of each $\n$ in
the field of its neighbors.

The  equation $\nabla^2 \n=0$ has
topological soliton solutions, or ``skyrmions''
where the mapping $\n(\r):{\bf R}^2\to S^2$ covers the sphere once as
the point $\r=(x^1,x^2)$ covers the plane. Using spherical polar
coordinates $\theta$, $\phi$ to parameterize $S^2$ these solutions can
be written
$$
e^{i\phi}\cot \theta/2=\frac a z,
\eqno(2.10)
$$
where $z=x^1+ix^2$. For anti-skyrmions we replace $z$ by its
conjugate.
The spins in (2.10) point up at the origin and gradually tilt down as
one moves outwards. They point straight down at infinity.

It is fairly easy to show [8] that a multi-skyrmion
solution to $\nabla^2 \n=0$ is given by
$$
e^{i\phi}\cot \theta/2 =f(z),
\eqno (2.11)
$$
where $f(z)$ is any rational function of $z$. The number of skyrmions
--- {\it i.e.\/} the degree or winding number of the map
${\bf R}^2\to S^2$ ---  is given by the number of poles (equivalently
by the number of zeros) of $f(z)$. For the simple stiffness term in
(2.6) the strain energy of these multi-skyrmion solutions is
independent of the parameters in the rational function. In particular
single skyrmions  of different scale $a$ are all degenerate in energy.
The skyrmions in the quantum Hall effect have a definite size
determined by a competition between coulomb repulsion  and the Zeeman
energy [1].  They are more compact than the solutions given by
(2.10).

Now we investigate the mobility of the skyrmions. Let us denote the
spin configuration of a skyrmion centered at the origin by $\n_0(\r)$.
We  introduce a collective coordinate $\R=(R^1,R^2)$ so that the spin
field of a moving skyrmion can be  written, at least as a first
approximation, as $\n(\r,t)=\n_0(\r-\R(t))$. We insert this ansatz
into (2.8) and see what action it costs to move the skyrmion round a
closed path in the plane.

We therefore wish to evaluate
$$
S=-\rho J\int \dot\n(\r,t)\cdot {\bf A}(\n(\r,t))\, dt\,d^2x,
\eqno(2.12)
$$
with $\n(\r,t)=\n_0(\r-\R(t))$. It is actually more convenient to
consider the variation of (2.12) under a small perturbation in the
path $\R(t)$. We then use
$$
\delta S=- \rho J \int  \n\cdot (\delta \n \wedge \dot \n)\,dt\,d^2x
\eqno (2.13)
$$
with
$$
\dot \n =- \frac{\partial}{\partial x^i}
\n_0(\r-\R)\,\dot R^i,
\eqno (2.14)
$$
and
$$
\delta \n = -
\frac{\partial}{\partial x^i}
\n_0(\r-\R)\,\delta R^i.
\eqno (2.15)
$$
Thus
$$
\eqalign{
\delta S=&-\rho J \int  \n_0(\r-\R)\cdot (\partial_i
\n_0(\r-\R)\wedge \partial_j\n_0(\r-\R))\delta R^i \dot R^j \,dt\,d^2x\cr
=&-\rho J \int \delta R^i\dot R^j \left \{\int
\n_0(\r-\R)\cdot(\partial_i\n_0(\r-\R)\wedge\partial_j\n_0(\r-\R))
\,d^2x\right\}\,dt.\cr
}
\eqno(2.16)
$$
The factor in braces in  (2.16) is independent of $R$ and is equal to
$4\pi \N \epsilon_{ij}$ where $\N$ is the degree of the map
$\n:{\bf R}^2\to S^2$ {\it i.e.\/} the skyrmion number. We therefore
have that
$$
\delta S=- 4\pi \N \rho J \oint (\delta R^1 \dot R^2-\delta R^2
\dot R^1)\,dt.
\eqno (2.17)
$$
This is the variation of
$$
S=-2\pi \N \rho J \oint ( R^1\dot R^2-R^2\dot R^1)\,dt.
\eqno (2.18)
$$
When used in the path integral $\Z=\int d[\n]\exp iS$, (2.18) means
that the skyrmion accumulates a phase of $2\pi$ for every
spin-$\frac 12$ it encircles. The physical consequence of this phase
may be  recognized by noting that the right hand side of (2.18) is
the same term that  occurs in the action for a particle of charge $\N$
moving in a uniform magnetic field of strength  $2\pi\rho J $. Like
the charged particle therefore,  the skyrmion must experience a
transverse  ``Lorentz'' or  ``Magnus'' force proportional to its
velocity. We will use the latter designation  because the force is
proportional to the surrounding spin density.

The Magnus force pins the skyrmion in place just as an electron in the
lowest Landau level is pinned in place by the quenching of its
kinetic energy.
 The addition to the lagrangian of an inertial term proportional to
$|\dot\n|^2$  would induce a mass for the  skyrmion and permit it to
make  cyclotron orbits, but in the absence of scattering its
wandering would still be strongly restricted.  A gradient in an
additional Zeeman term will however lead to the skyrmion drifting
along the skyrmion's Zeeman-energy contours at such a speed that the
Magnus force balances the Zeeman-energy gradient.

At first sight it is surprising that we accumulate a  phase
proportional to the area encircled by the skyrmion. The phase comes
from a term involving the space integral of $\dot \n$, and far from
the skyrmion the spins do not move. Naively this  would lead us to
expect a phase proportional  to at most the {\it length\/} of the
skyrmion trajectory, and not to the {\it area\/} it encloses. The
incipient paradox is resolved by keeping track of the motion of any
particular spin due to a  family of skyrmion trajectories. Assume the
skyrmion first passes our chosen spin on its right, but then the
trajectories gradually sweep across the spin so that the skyrmion
eventually passes the spin on its left. We see that the corresponding
family of spin paths on $S^2$ start with a small loop near the south
pole, then, as the skyrmion passes closer to the spin, the
loop in $S^2$ grows in area, circling the equator of the sphere when
the center of the skyrmion goes exactly through the spin. As the
skyrmion  passes further to the left the loop begins to shrink
towards the south pole again, but the area enclosed by the loop
continues to grow, reaching $4\pi$, or the entire sphere, when the
skyrmion passes far to the left.  Thus, although the spins far to the
left of the skyrmion's trajectory hardly move, they must be counted
as contributing $4\pi$ to $\oint \dot \n\cdot {\bf A}\,dt$.

\vskip 12 pt

\line{{\bf 3. Quantum Hall effect}\hfil}

We now wish to examine the dynamics of the skyrmions in the quantum
Hall effect. We will describe this system by means of the
Zhang-Hansson-Kivelson (ZHK) model [9], as modified by Kane and Lee
[10] to take into account the effects of spin.
This is one of the
models that
was used by Sondhi {\it et al.\/} [1].

We will take as lagrangian for this system
$$
\eqalign{
L=i\phi^\dagger(\partial_0-&i(a_0+eA_0))\phi-\frac 1{2m^*}
|(\partial_i-i(a_i+eA_i))\phi|^2 \cr
-&\frac \lambda2 (|\phi|^2-\rho_0)^2+
\frac 1{4\Theta}\epsilon^{\mu\nu\sigma}a_\mu\partial_\nu
 a_\sigma.\cr
}
\eqno(3.1)
$$
Here $\phi=(\phi_1,\phi_2)$ is a two-component complex scalar field.
The quantity $\Theta$ is the statistics parameter which must take one
of the values $\Theta=(2n+1)\pi$ in order that the boson field $\phi$
represents a fermion. $A_\mu$ is the external
electromagnetic field and  $m^*$ is the effective mass of the electron.
Repeated roman indices imply sums over the spatial directions
$i=1,2$, while repeated greek indices imply sums over both space and
time directions $\mu=0,1,2$.

We should also include Coulomb and Zeeman interactions. These are
essential for determining the energy and size of  the skyrmions, but
they  are not important ingredients in the topological effects we are
studying here. We will therefore omit them from our expressions so as
not to unduly clutter our formulae.

In order to find a solution with uniform density
$\rho \equiv(|\phi_1|^2+|\phi_2|^2)=\rho_0$ at filling
fraction $\nu=1/|2n+1|$ we must adjust the magnetic field
$B_z=\partial_1A_2-\partial_2A_1$ so that $eB_z=2\Theta \rho_0$, with
the sign of $\Theta=(2n+1)\pi$ chosen so as to ensure $\rho_0>0$. At
this magic value the Chern-Simons field cancels the effects of the
external magnetic field on $\phi$. We assume that this adjustment has
been made, and that the solution selected by the Zeeman term has
$\phi_1=\sqrt{\rho_0}$, $\phi_2=0$.  We now examine the dynamics  of
the fluctuations about this uniform solution
under the conditions that the density varies sufficiently slowly that
derivatives of $\rho$ can be ignored. (Including the derivatives of
$\rho$ merely adds a ``quantum pressure'' of the kind familiar from
the Gross-Pitaevski model of a quantum fluid.) We therefore replace
(3.1) by a lagrangian reminiscent of the $CP^1$ version of the
non-linear $\sigma$-model [10]
$$
\eqalign{
L=i\rho {\bf z}^\dagger(\partial_0-&i(a_0+eA_0)){\bf z}-
\frac {\rho}{2m^*}
|(\partial_i-i(a_i+eA_i)){\bf z}|^2 \cr
-&\frac \lambda 2(\rho-\rho_0)^2+
\frac 1{4\Theta}\epsilon^{\mu\nu\sigma}a_\mu\partial_\nu
a_\sigma.\cr
}
\eqno(3.2)
$$
Here ${\bf z}$ is a two component complex field with the
constraint ${\bf z}^\dagger {\bf z}=|z_1|^2+|z_2|^2=1$. We can now
partially decouple the spin dynamics from the charge transport by
using  the identity
$$
\frac {\rho}{2m^*}|(\partial_i-i(a_i+eA_i)){\bf z}|^2
=\frac {\rho}{8m^*}(\nabla\n)^2+
\frac {m^*}{2\rho}{\bf J}^2,
\eqno (3.3)
$$
where  ${\bf J}\equiv(J^1, J^2)$ with
$$
J^i=\frac{\rho}{m^*i}({\bf z}^\dagger \partial_i {\bf z}
-i(a_i+eA_i))
\eqno (3.4)
$$
is the number current, $n^a={\bf z}^\dagger \sigma^a {\bf z}$
is the local spin
direction, and  $\nabla\n\equiv(\partial_1\n,\partial_2\n)$.

Our lagrangian now looks like
$$
L=i{\rho}({\bf z}^\dagger\partial_0 {\bf z}
-i(a_0+eA_0))-\frac {\rho}{8m^*}(\nabla\n)^2-
\frac {m^*}{2\rho}{\bf J}^2  +\ldots
\eqno (3.5)
$$
where the dots represent the potential and
Chern-Simons terms that are temporarily uninteresting.

 We next demonstrate that (3.5) represents the action for a
ferromagnet of the kind considered in section 2.
We already have a spin stiffness term with $K={\rho}/4m^*$, and we will now
show that the time derivatives of $z$ provide the
$\dot\n\cdot {\bf A}(n)$ term.

The expression appearing as part of the currents and in the kinetic
term, $K_\mu \equiv {\bf z}^\dagger\partial_\mu {\bf z} $ is
essentially the Berry connection on $S^2$. It is no surprise, then,
that $K_\mu$ satisfies
$$
\partial_\mu K_\nu-\partial_\nu K_\mu=\frac {i}{2}
\n\cdot(\partial_\mu\n\wedge\partial_\nu\n).
\eqno (3.6)
$$
We can use this equation to find the variation of $K_\mu$ under a
variation of the direction $\n$. Replacing one of the partial
derivatives by $\delta$ gives us
$$
\delta K_\mu=\frac {i}{2}\n\cdot(\delta \n \wedge \partial_\mu \n)
+\partial_\mu \Lambda,
\eqno (3.7)
$$
where $\Lambda={\bf z}^\dagger \delta {\bf z}$.
To be rather  more concrete we could write
${\bf z}=U{\bf z}_0$ where ${\bf z}_0=\left(\matrix{1\cr 0\cr}\right)$
 and $U\in SU(2)$. We
make a variation in
 $U$ so that $\delta U U^{-1}=i \delta {\bf w}\cdot\sigma/2$ and find that
 $$
\eqalign{
\delta \n=& \n\wedge \delta{\bf w}\cr
\delta K_\mu=&-\frac {i}{2}\delta{\bf w}\cdot \partial_\mu \n +
\frac{i}{2}\partial_\mu (\n\cdot\delta{\bf w}).\cr
}
\eqno (3.8)
$$
However we make the variation, the second term in $\delta K_\mu$ is a
pure gauge transformation, and  so has no effect on the lagrangian
provided the number conservation equation $\dot \rho+
\nabla \cdot {\bf J}$ is satisfied.
Substituting the variation from (3.8) into (3.5) (and again ignoring
gradients of $\rho$) gives us
$$
\frac {\rho}{2}(\partial_t+{\bf v}\cdot \nabla)\n
=\frac {\rho}{4m^*}\n\wedge (\nabla^2\n),
\eqno (3.9)
$$
where ${\bf v}={\bf J}/\rho$ is the local electron-fluid velocity.

We see that (3.9) is the Landau-Lifshitz equation for a density
$\rho$ of spin-$\frac 12$ particles ---  except that  the
time derivative has been replaced by  a convective derivative. (The
$({\bf v}\cdot\nabla)\n$ term comes from the variation of ${\bf J}^2$ in
(3.5).)
This modification is to be expected because the  galilean invariance
implicit in (3.1) requires the spin waves to be
carried along with the local flow.

Provided the electron fluid remains stationary, the spins in the
quantum Hall system obey exactly the same equation as in a
conventional ferromagnet. The solitons, although they are
electrically charged, must therefore experience  only the same
Magnus force that they feel in the neutral ferromagnet.  Now the
skyrmion does carry its charge along with it so a  small velocity
field is induced by the skyrmion motion ---  but this does not affect
our conclusions as we will now show by making a partial duality
transformation.

We first promote the current, ${\bf J}$, to the status of an
independent dynamical variable  by making a Hubbard-Stratanovich
transformation.    The lagrangian becomes
$$
\eqalign{
L=i\rho({\bf z}^\dagger\partial_0{\bf z}-&i(a_0+eA_0))
+i({\bf z}^\dagger\partial_i{\bf z}-i(a_i+eA_i))J^i
+\frac {m^*}{2\rho}{\bf J}^2 -\frac \lambda 2(\rho-\rho_0)^2\cr
+&\frac {1}{4\theta}\epsilon^{\mu\nu\sigma}a_\mu\partial_\nu a_\sigma
-\frac {\rho}{8m^*}(\nabla \n)^2.\cr
}
\eqno(3.10)
$$
Integrating over the $U(1)$ phase degree of freedom in ${\bf z}$
enforces the current conservation law as a constraint, so we can write
the current/density three-vector $(\rho\equiv J^0, J^1,J^2)$ as the
curl of a three-dimensional vector field.
We set $J^\mu_{\{0\}}\equiv(\rho_0, 0,0)$ equal to
$\epsilon^{\mu\nu\sigma}\partial_\nu\A^{\{0\}}_\sigma$
 and
$$
J^\mu-J_{\{0\}}^\mu =\epsilon^{\mu\nu\sigma}\partial_\nu\A_\sigma.
\eqno (3.11)
$$

At this point we  also integrate out the  Chern-Simons field $a_\mu$.
After some further integration by parts and use of the relation
$2\Theta \rho_0=eB_z$ we find
$$
L=2\pi \J^\mu(\A_\mu+\A_\mu^{\{0\}})
-\Theta J^\mu \A_\mu +\frac {m^*}{2\rho}{\bf J}^2-
\frac \lambda 2(J^0-J^0_{\{0\}})^2 -\frac{\rho}{8m^*}(\nabla \n)^2,
\eqno (3.12)
$$
where
$$
\J^\mu=
\frac {1}{2\pi i}\epsilon^{\mu\nu\sigma}\partial_\mu \z_\alpha \partial_\nu
z_\alpha
\eqno (3.13)
$$
is the skyrmion number current.

We can make the physics content of (3.12) clearer  by first
linearizing by
setting $\rho=\rho_0 $ in the kinetic energy term. Then, for cosmetic
reasons, we  adjust the units of length and time  so that
$c\equiv \sqrt{\lambda\rho_0/m^*}$,
the velocity of density waves in the absence of the magnetic field,
becomes unity, and define the field strength tensor
$\F_{\mu\nu}=\partial_\mu\A_\nu-\partial_\nu \A_\mu$ to be the dual of
the electron number current.

With these changes we have
$$
L=2\pi \J^\mu(\A_\mu+\A_\mu^{\{0\}}) -\frac 12
\Theta \epsilon^{\mu\nu\sigma}\A_\mu \F_{\mu\nu}-\frac \lambda 4
\F_{\mu\nu}\F^{\mu\nu}-\frac{1}{8\lambda}(\nabla\n)^2.
\eqno (3.14)
$$
We see that the skyrmion number current acts a source for a
topologically massive gauge field $\A_\mu$ [11], and also sees the
background field $\A^{\{0\}}$. The curl of $\A^{\{0\}}$ is equal   to
both $\rho_0$ and $eB_z/2\Theta$, so the interaction with the
background field provides a phase for each skyrmion world line that
can be interpreted as producing either the Magnus force on the
skyrmion, or, provided that the skyrmion number current and the
electron number current associated with the skyrmion world line are
proportional with proportionality factor $\Theta/\pi$, the Lorentz force.
 The interaction with the topologically massive gauge field
 provides a phase factor  that depends on all skyrmions present and
gives the skyrmion  Fermi (for $\nu=1$), or anyon (for
$\nu=\frac {1}{|2n+1|}$) statistics.  The topologically massive
gauge field also has its own degrees of freedom. These are the   gapped
magneto-phonons or magneto-plasmon density fluctuation modes.

If we momentarily freeze out the density fluctuation modes by
ignoring the
$\F_{\mu\nu}\F^{\mu\nu}$
term in (3.14),
the equation of motion from varying $\A_\mu$  is
$$
2\pi \J^\mu=\Theta \epsilon^{\mu\nu\sigma}\F_{\nu\sigma}=2\Theta J^\mu.
\eqno (3.15)
$$
This shows  that, up to the expected factor that reflects the skyrmion
possessing total charge $\frac 1{2n+1}$, the topological current
density  coincides with the electron current density. If we  now
reintroduce the $\F_{\mu\nu}\F^{\mu\nu}$  term the two current
densities no longer exactly coincide, but, because the extra term
that appears in the equation of motion is a pure divergence, the
{\it total\/} currents flowing along a skyrmion world line are the
same. This confirms that the Magnus force and Lorentz force on the
skyrmion are indeed equal, and that they are merely different
physical interpretations of the same phase factor.

\vskip 12pt
\line{\bf 4. Discussion \hfil}

The same disappearing-force  phenomenon occurs for the
vortex-like charged quasiparticles in the original ZHK model.   These
 quasiparticles also look as if they should experience a Magnus force,
both because of the phase winding in the order parameter and because
of there is circulation around the vortex. They should also be acted on by
a Lorentz force because of their charge.  Once again only one force
survives [12]. In this case however the effect is not so startling.
The circulation vanishes at large distance from the quasiparticles
and, anyway, they  are very much children of the quantum Hall
phase and one is less likely to grant them attributes outside it.

Skyrmions  on the other hand have an independent existence in planar
ferromagnets. They experience a Magnus force when neutral and so
would naively be expected to experience an additional Lorentz force
when given a charge. It requires an intricate conspiracy for the two
forces to become one. Such intricacies are not unknown in
the ferromagnets. Identifying the force on a soliton is equivalent to
identifying the momentum that the force is changing. For the
Landau-Lifshitz system this is  non-trivial. In the continuum
approximation to a  ferromagnet with spins in fixed locations the
momentum operator $\hat P$ itself is not well-defined. Only the
translation operator $T_a=\exp \{i a \hat P\} $
makes sense, and then only for translations through a distance $a$
that takes one past  an integer number of $J=\frac 12$ spins
[13]. When the spins are attached to mobile electrons, as they are in
the Hall effect, the system is manifestly translation invariant and
must therefore have a well-defined momentum.  There is, however, no
gauge invariant separation between the momentum residing in
 the collective orientational order of the spins and the motion of
the electrons
 [14].

In [14] Volovik pointed out that, once a gauge tranformation has been
applied to align the local spin quantization frame along $\n$, the
electrons respond to the  winding-number density of $\n$ as if it
were a magnetic flux. This observation  provides a simple picture of
what is happening here. A moving flux produces an electric field and
in Volovik's one-dimensional example the passage of a skyrmion
accelerates the electrons, producing a spectral flow that effectively
transfers a single electron from one side of the Fermi surface to the
other. His electrons have thus gained momentum $2k_f=2\pi \rho$.

A two-dimensional version of this phenomenon is the physical origin of
the Magnus force.  For a  ferromagnet composed of mobile electrons
with an ungapped fermi-surface the electrons gain both energy and
momentum from the moving skyrmion. A force must be applied to the
skyrmion to provide this momentum and energy. When the  electron gas
is in a quantum Hall phase, however, no change in occupation number
through spectral flow is possible because there is a gap in the
spectrum.  The skyrmion is therefore no longer able to excite the
system --- but the moving ``flux'' is still attempting to transfer
mechanical momentum to the electron fluid. The electrons are unable
to accelerate in response because they are  locked in place  by the
external magnetic field. The mechanical momentum generated by the
changing ``flux'' must therefore be transferred to the  magnet that
creates this field. Now exerting a force on the magnet requires the
electrons to produce a {\it real\/} magnetic field. Fortunately the
Hall effect itself requires the geometric ``flux'' to accumulate
extra charge in its vicinity and the electrons are able to  generate
the necessary field from the current created by dragging this charge
along with  the skyrmion. Since magnets obey Newton's third law, the
magnet then  produces an equal and opposite force on the skyrmion.

 By this sequence of maneuvers the Hall effect electrons have
transformed the original  Magnus force into a Lorentz force.

\vskip 12pt

\noindent
{\bf Note added:} Since finishing this manuscript I became aware that
the Magnus force is briefly discussed in a recent paper by Read and
Sachdev [15].

\vskip 12pt
\line{\bf Acknowledgements \hfil}

The question addressed in this paper arose in a discussion with Amir
Caldeira. Most of the work was carried out at the Erwin
Schr{\"o}dinger Institute in Vienna and I would
like to the thank the staff and members of the ESI for their
hospitality. I must also thank Eduardo Fradkin and  Frank Gaitan for
discussions about both the physics and the text.  This work was
supported by the National Science Foundation  under grant
DMR94-24511.

\vskip 12pt
\line{{\bf References}\hfil}
\item{[1]} S.~L.~Sondhi, A.~Karlhede, S.~A.~Kivelson, E.~H.~Rezayi,
Phys.~Rev. {\bf B47} (1993) 16419.

\item{[2]} S.~E.~Barrett, G.~Dabbagh, L.~N.~Pfeiffer, W.~W.~West,
R.~Tycko, Phys.~Rev.~Lett. {\bf 74} (1995) 5122;
R.~Tycko, S.~E.~Barrett, G.~Dabbagh, L.~N.~Pfeiffer, K.~W.~West,
Science, {\bf  268} (1995) 1460.

\item{[3]} J.~R.~Klauder, Phys.~Rev.~{\bf D 19} (1979) 2349.

\item{[4]} For a recent  account see: E.~A.~Kochetov,
J.~Math.~Phys.~{\bf 36} (1995) 4667.

\item{[5]} M.~Stone, Nucl.~Phys. {\bf B314} (1989) 557

\item{[6]} A.~A.~Kirillov, {\it Elements of the Theory of
       Representations\/}, (Springer-Verlag 1976).

\item{[7]} L.~D.~Landau, E.~M.~Lifshitz, Phys.~Z.~Sowjet, {\bf 8} (1935)
       3470; W.~D{\"o}ring, Z.~Phys {\bf 124} (1947) 501; C.~Herring,
       Ch.~Kittel. Phys.~Rev., {\bf 81} (1951) 869.

\item{[8]} For a review see R. Rajaraman {\it Solitons and
       Instantons\/}, (North Holland, Amsterdam 1982 )

\item{[9]} S.-C.~Zhang, T.~H.~Hansson, S.~A.~Kivelson,
       Phys.~Rev.~Lett., {\bf 62} (1989) 82.

\item{[10]} D.~H.~Lee, C.~L.~Kane, Phys.~Rev.~Lett. {\bf 64} (1990) 1313.

\item{[11]} J.~Schonfield, Nucl.~Phys. {\bf B 185} (1981) 157; S.~Deser,
 R.~Jackiw, S.~Templeton, Annals of Physics (NY) {\bf 140}(1982) 372

\item{[12]} M.~Stone, Phys.~Rev. {\bf B42} (1990) 212.

\item{[13]} F.~D.~M.~Haldane, Phys.~Rev.~Lett. {\bf 57} (1986) 1488.

\item{[14]} G.~E.~Volovik,  J. Phys {\bf C20} (1987) L83

\item{[15]} N.~Read, S.~Sachdev, Phys.~Rev.~Lett. {\bf 75} (1995)
       3509.

\bye